\documentclass[14pt]{article}

\usepackage[T2A]{fontenc}
\usepackage{graphicx}
\usepackage{amsmath}
\usepackage{ulem}
\usepackage[english]{babel}

\begin{document}

\title{Comparison of pulsar parameters for some anomalous types}
\author{I.F.Malov\footnote{%
                  malov@prao.ru } [1],    
        H.P.Marozava\footnote{%
                  ganna.marozava@gmail.com} [2]}  
\date{}
\maketitle
[1] P.N.Lebedev Physical Institute of the Russian Academy of Sciences

[2] Pushchino State Institute of Natural Sciences

\abstract{The comparative analysis of parameters is carried out for two samples of radio pulsars. Objects of the first sample have periods P > 2 sec, the second is characterized by magnetic fields at the neutron star surface $Bs > 4.4\times10^{13}$ G. The main goal of this analysis is to find some differences between normal pulsars and anomalous X-ray pulsars (AXP) and soft gamma-ray repeaters (SGR) with the similar values of periods and magnetic fields. It is shown that the dependence of period derivatives on periods for pulsars with long periods is $dP/dt \propto P^{1.67}$. This means that we must search for another braking mechanism differing from the magnetic dipole one which gives $dP/dt \propto P^{-1}$ or use some specific distributions of parameters. On the other hand, radio pulsars with strong magnetic fields and AXP/SGR can be described by the magnetic dipole model. Radio pulsars with P > 2 sec have ages on average two orders of magnitude higher than objects with $Bs > 4.4 \times 10^{13}$ G and are located much higher above the Galactic plane. The obtained results show that the main feature of AXP/SGR is the strong magnetic field and not the large period.}

\section{Introduction}

During the last two decades intense investigations of anomalous X-ray pulsars (AXP) and soft gamma-ray repeaters (SGR) were carried out. The nature of these objects is still unclear. The magnetar model put forward in 1992 \cite{Dunkan} is the most popular at the moment. As alternatives some other models were considered, namely, white dwarfs with strong magnetic fields \cite{Paczynski, Usov1993}, an accretion from the surrounding disc \cite{Trumper, Bisnovatiy}, the drift model \cite{Malov2003} suggesting the existence of drift waves at the periphery of the magnetosphere. These waves can cause a periodic modulation of outgoing radiation. There is a number of publications considering some processes at the surface of a strange star \cite{Dar, Alcock, Usov2001, Xu}. Finally, it is proposed to described all anomalous pulsars and AXP/SGR including as the evolutionary sequence of states of the degenerate Kepler disc formed during the explosion of quark-nova \cite{Ouyed2002, Ouyed2002-1, Ouyed2002-2}. All the proposed models are not without difficulties, and the question of an adequate theoretical description of the observational data remains open.

AXPs and SGRs can be characterized by long periods, intervals between successive pulses, and strong magnetic fields calculated using the formula

\begin{equation}
Bp = 6.4 {\times}  10^{19} (P dP/dt)^{1/2} ,
 \label{eq:ref}
\end{equation}

 obtained in the frame of the magnetic dipole braking model
 
\begin{equation}
-I\Omega d\Omega/dt = \frac{ Bp  R_*^6 \Omega ^{4} Sin^2 \beta}{ 6c^3  }
 \label{eq:ref}
\end{equation}

Here I is the moment of inertia of a neutron star, $\Omega = 2\pi/P$ is the angular velocity of its rotation,  Bp  is the polar magnetic field, R$_*$ is the radius of the neutron star, $\beta$ is the angle between the magnetic moment and the rotation axis, с is the speed of light. The ATNF catalog \cite{Manchester} contains values of equatorial magnetic fields. They are two times smaller than polar ones. Moreover it was suggested there that $Sin\beta = 1$ for all pulsars. Corrections to values of magnetic fields given in the ATNF catalog have been considered by Nikitina \& Malov in \cite{Nikitina}.

The renewable catalog of AXP/SGR \cite{mcgill} gives for these objects P > 2 sec, $Bp \geq  Bsh = 4.4 \times 10^{13} G$. In \cite{Olausen} the distributions of a number of AXP/SGR parameters were compared with the corresponding distributions for the main population of pulsars from the ATNF catalog. Here, we propose to analyze the difference in the distributions of the AXP/SGR parameters and “normal” radio pulsars with periods P > 2 sec and magnetic fields at the pole $Bp > 4.4\times 10^{13} G$ to understand the difference between quiet neutron stars with these parameters and stars with a flare activity. Two points should be noted.

1) The PSR J1846-0258, object with P = 0.33 sec included in the catalog \cite{mcgill} exhibits some outburst activity only occasionally and so far should be considered as an unconfirmed candidate for AXP, therefore we excluded it from the further analysis.

2) The catalog \cite{Manchester} lists magnetic fields at the equator of a neutron star, which are two times weaker than fields at its poles. In this regard, we consider further pulsars with catalog fields $B > 2.2  \times 10^{13} G$. For these objects, the polar field must be higher than the Schwinger field ($Bsh = 4.4\times 10^{13} G$).

\section{Features of quiet pulsars with long periods}

First of all, let’s analyze the difference between “normal” radio pulsars with periods P > 2 sec and P < 2 sec. According to the predictions of Sturrock \cite{Sturrock}, long-period pulsars (P > 1 sec) should suppress the cascade production of the electron-positron secondary plasma, and as a result their characteristics should differ from the parameters of pulsars with shorter periods. In this section we compare the characteristics of two samples of radio pulsars, for the first of them 0.1 sec < P < 2 sec, for the second P > 2 sec. The first sample does not include objects with the shortest periods ($P < 0.1 sec$), which differ in a number of parameters. All millisecond (“recycled”) pulsars belong to this population (see \cite{Loginov2013, Loginov2014}). In addition, pulsars in binary systems and globular clusters are excluded from the consideration, since their features can be associated with the influence of companions and surrounding stars.

Let us consider the distribution of period derivatives in the samples mentioned.

\includegraphics[width=10cm]{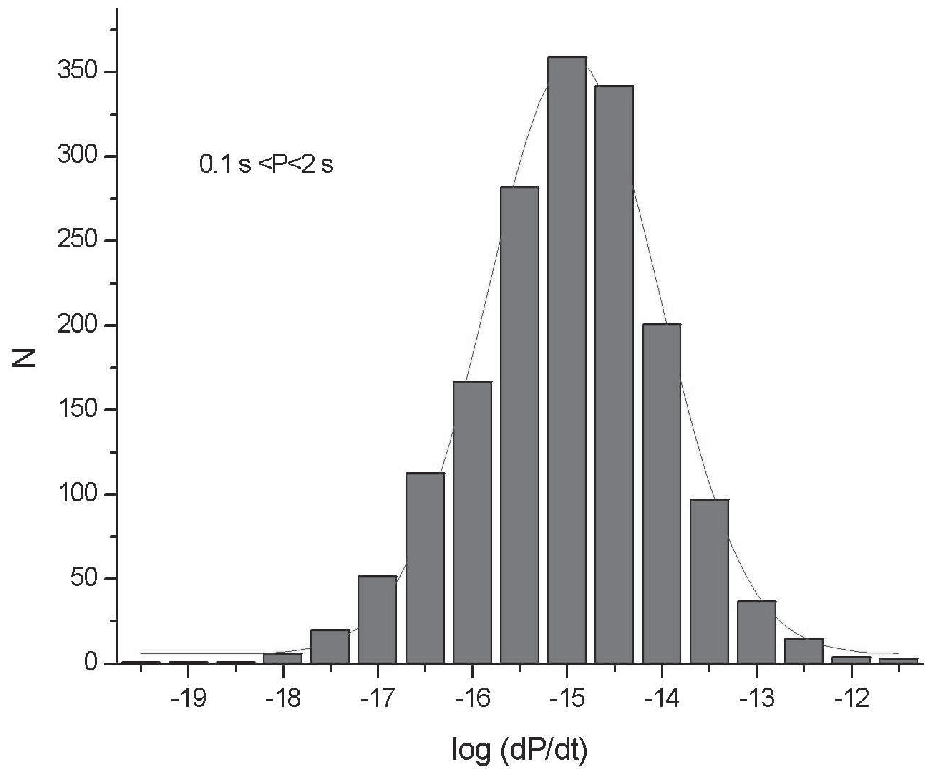}

\includegraphics[width=10cm]{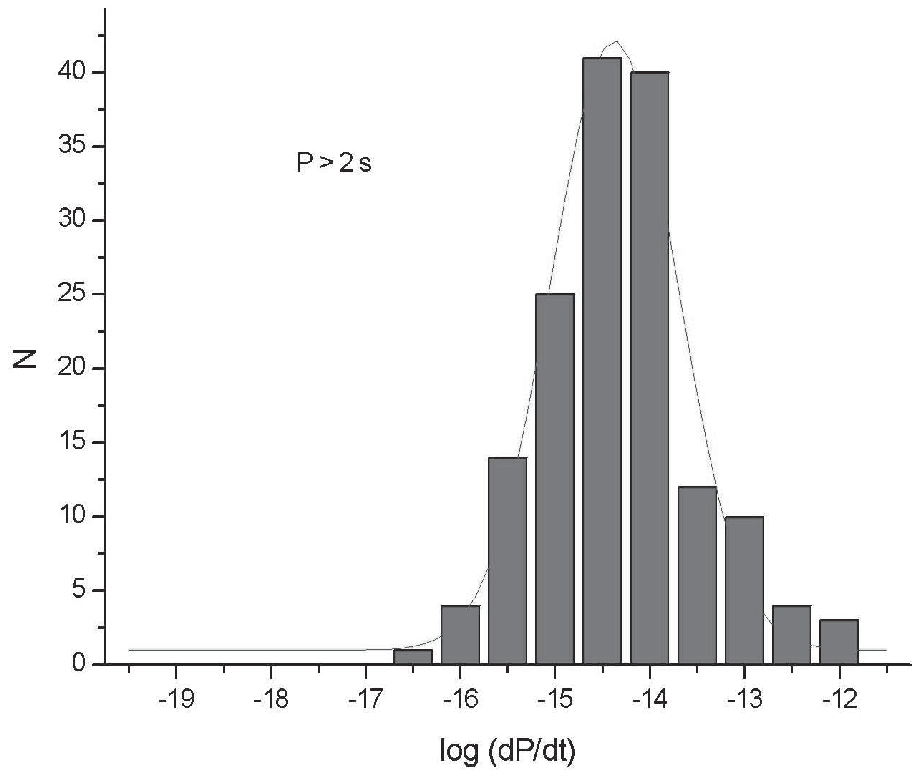}

Fig. 1 - Distributions of period derivatives for the two considered pulsar samples
\vspace{\baselineskip}

Figure 1 shows the histograms of N(dP/dt) together with the parameters of the Gaussians inscribed in them:

\begin{equation}
N = 352 exp \left[- \left( \frac{-dP/dt -14.94}{1.81} \right)^2\right] ,  \qquad  (0.1 sec < P < 2 sec)
 \label{eq:ref}
 \end{equation}
 \begin{equation}
 N = 41 exp \left[- \left( \frac{-dP/dt -14.38}{1.33} \right)^2\right] , \qquad \qquad \qquad   (P > 2 sec) 
 \label{eq:ref}
\end{equation}

As follows from this figure, pulsars with long periods slow down, on average several times faster (median values of logdP/dt are equal to -14.0 and -14.9, respectively). These two histograms differ with the probability more than 99\% according to the Kolmogorov-Smirnov criterion.  Another feature of long-period pulsars is their dependence dP/dt(P) (Fig. 2), which indicates the inapplicability of the magnetic dipole braking model to this population. Indeed, a straight line drawn by the least squares method can be described by the equation:

\begin{equation}
log (dP/dt) = (1.67 \pm 0.32)log P - 14.92 \pm0.17
 \label{eq:ref}
\end{equation}

with the  correlation coefficient K = 0.40 and the probability of random distribution p <10$^{-4}$.
\vspace{\baselineskip}

\includegraphics[width=9.5cm]{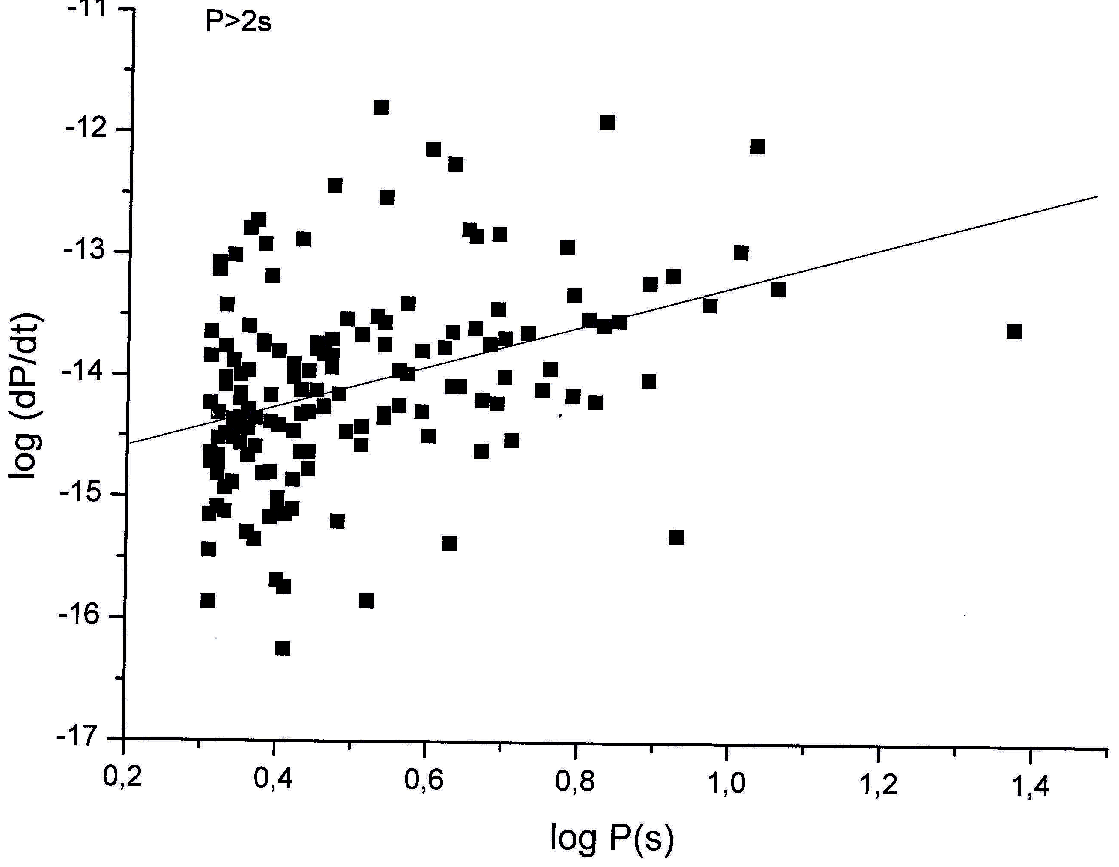}

Fig. 2 - Dependence of the period derivative on the period for 150 radio pulsars with P > 2 sec

\includegraphics[width=11cm]{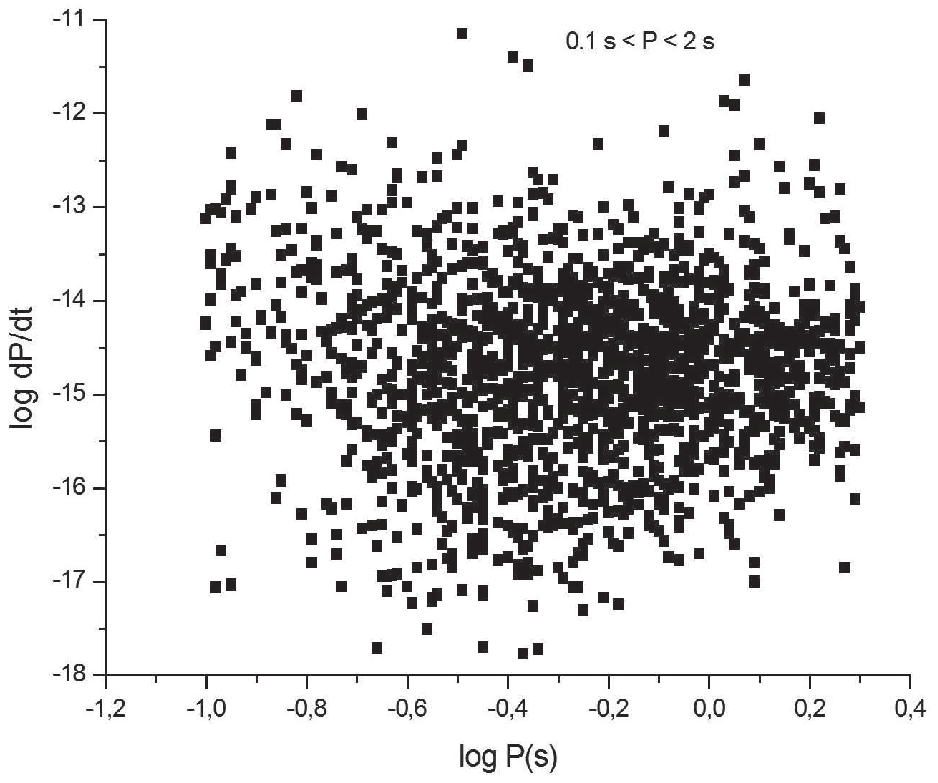}

Fig. 3 - Diagram dP/dt (P) for radio pulsars with periods from 0.1 to 2 sec
\vspace{\baselineskip}

This dependence differs radically from that predicted by relation (1) with a constant magnetic field ($P \propto P^{-1}$). It should be noted that dependence (5) is statistical, but it is difficult to imagine how it can be obtained by adding up individual magnetic dipole braking paths for individual objects. As for pulsars with short periods, there is no pronounced correlation dP/dt (P) for them (Fig. 3). Nevertheless, it is difficult to find any evidences of a magnetic dipole model here also. It can be concluded that this model as the main one for the samples under consideration is not applicable, and other braking mechanisms must be taken into account. Another explanation of this relationship may be find by the search for specific parameter distributions. Possible alternative processes leading to the pulsar braking are considered, for example, in ~\cite{Malov2017}. As a consequence of what has been said, the values of the magnetic field inductions given in the catalog ~\cite{Manchester} need to be corrected. However, there are no more accurate estimates currently. We will continue to operate with catalog fields, assuming that a distortion of their values occurs in a similar way for various pulsars and, if the absolute values are inaccurate, their relative values can be used to compare the samples under consideration.
Figure 4 shows the distribution of magnetic fields on the surface of the pulsars, which differ in the measured periods together with the Gaussians inscribed in them:

\begin{equation}
N = 724 exp \left[- \left( \frac{logBs -12.05}{0.90} \right)^2\right] ,  \qquad  (0.1 sec < P < 2 sec)
 \label{eq:ref}
 \end{equation}
 \begin{equation}
 N = 67 exp \left[- \left( \frac{logBs -12.93}{0.88} \right)^2\right] , \qquad \qquad \qquad   (P > 2 sec) 
 \label{eq:ref}
\end{equation}

\includegraphics[width=10cm]{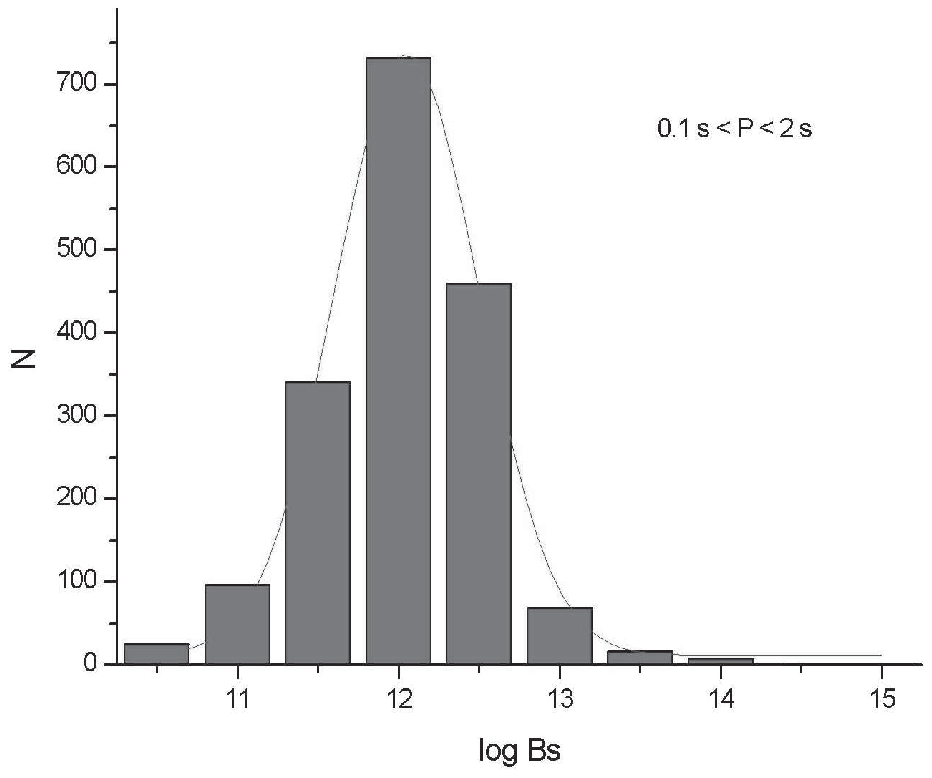}

\includegraphics[width=10cm]{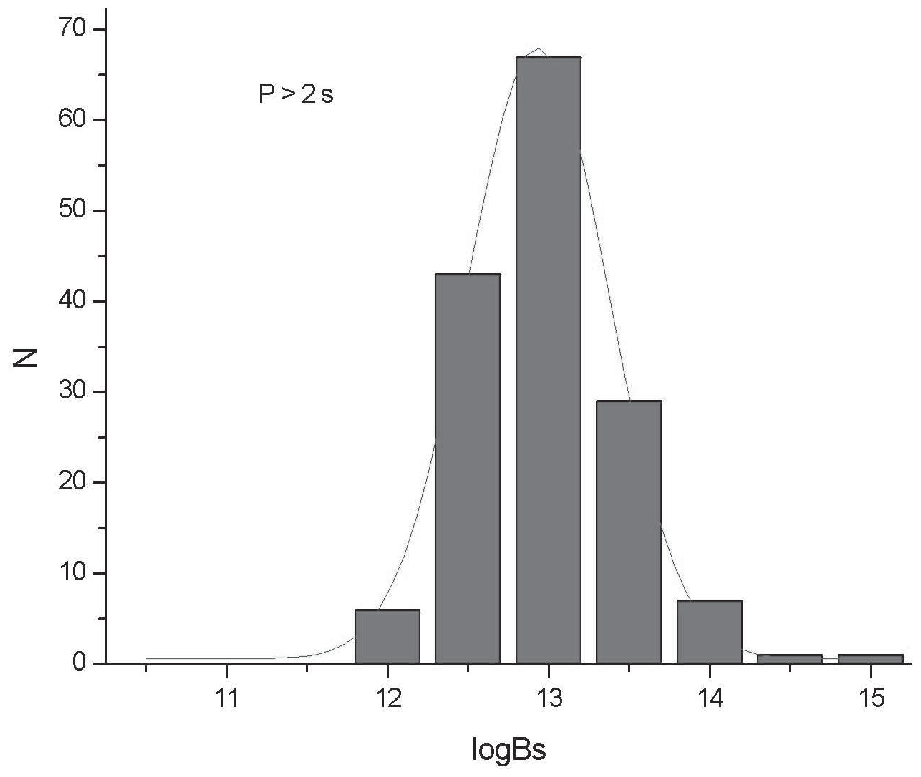}

Fig. 4 - Distributions of magnetic fields on the surface of pulsars with periods of 0.1 sec < P < 2 sec and  P > 2 sec
\vspace{\baselineskip}

As follows from these distributions, Bs for pulsars with longer periods are approximately an order of magnitude higher (the median values of their logarithms are 12.76 and 11.95 respectively). According to the Kolmogorov-Smirnov criterion, the difference is significant with a probability higher than 0.999. This result was to be expected, since for large values of P and dP/dt in equation (1) the value of B should be higher. The same can be said about the distributions of N(B$_{LC}$), N(dE/dt) and N(W$_{10}$) (Fig.5-7), since for a dipole field

\begin{equation}
B_{LC} = B_s  \frac{R_*^3}{R_{LC}^3} =  \frac{8\pi ^3 R_*^3}{c^3P^3},  \qquad  
 \label{eq:ref}
 \end{equation}
 
 where
 
 \begin{equation}
 R_{LC} = \frac{c}{\Omega} 
 \label{eq:ref}
\end{equation}

is the light cylinder radius, and

\begin{equation}
dE/dt = \frac{4\pi^2 I dP/dt}{P^3}  
 \label{eq:ref}
\end{equation}

\vspace{\baselineskip}

\includegraphics[width=10cm]{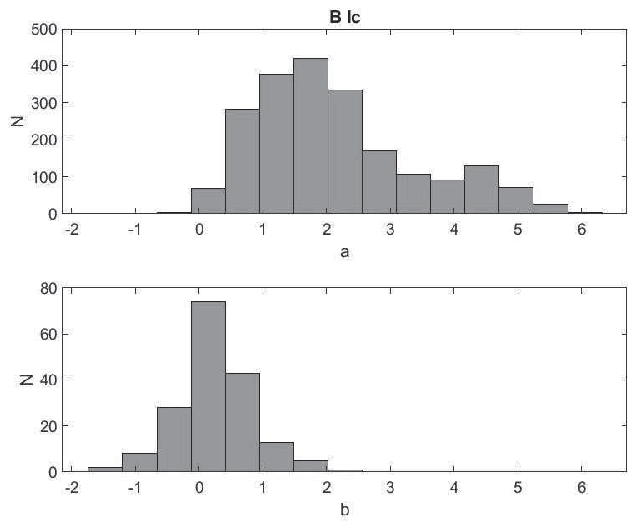}

Fig. 5 - Distributions of magnetic fields on the  light cylinder in pulsars with P < 2 sec (a) and P > 2 sec (b)
\vspace{\baselineskip}

\includegraphics[width=10cm]{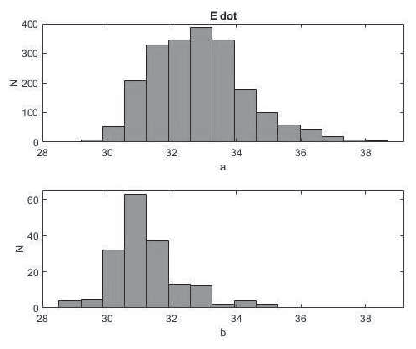}

Fig. 6 - Distribution of rotational energy loss rates. Pulsars with P < 2 sec (a) and with P > 2 sec (b)
\vspace{\baselineskip}

\includegraphics[width=10cm]{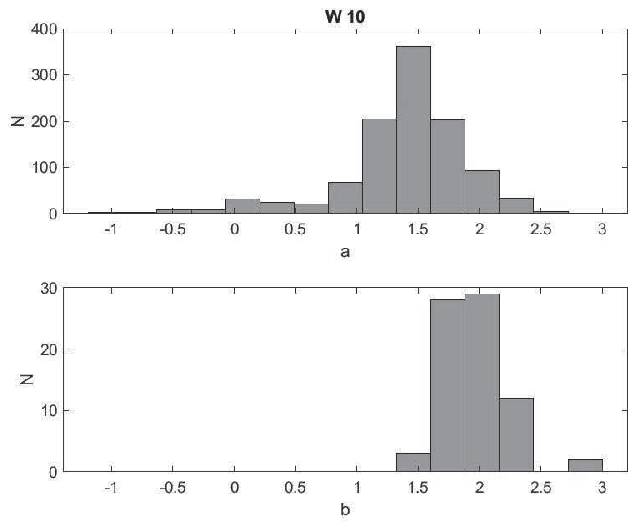}

Fig. 7 - Pulse width distributions for pulsars with P < 2 sec (a) and P > 2 sec (b). The logarithms of the widths, measured in milliseconds are shown in the abscissas.
\vspace{\baselineskip}

\includegraphics[width=10cm]{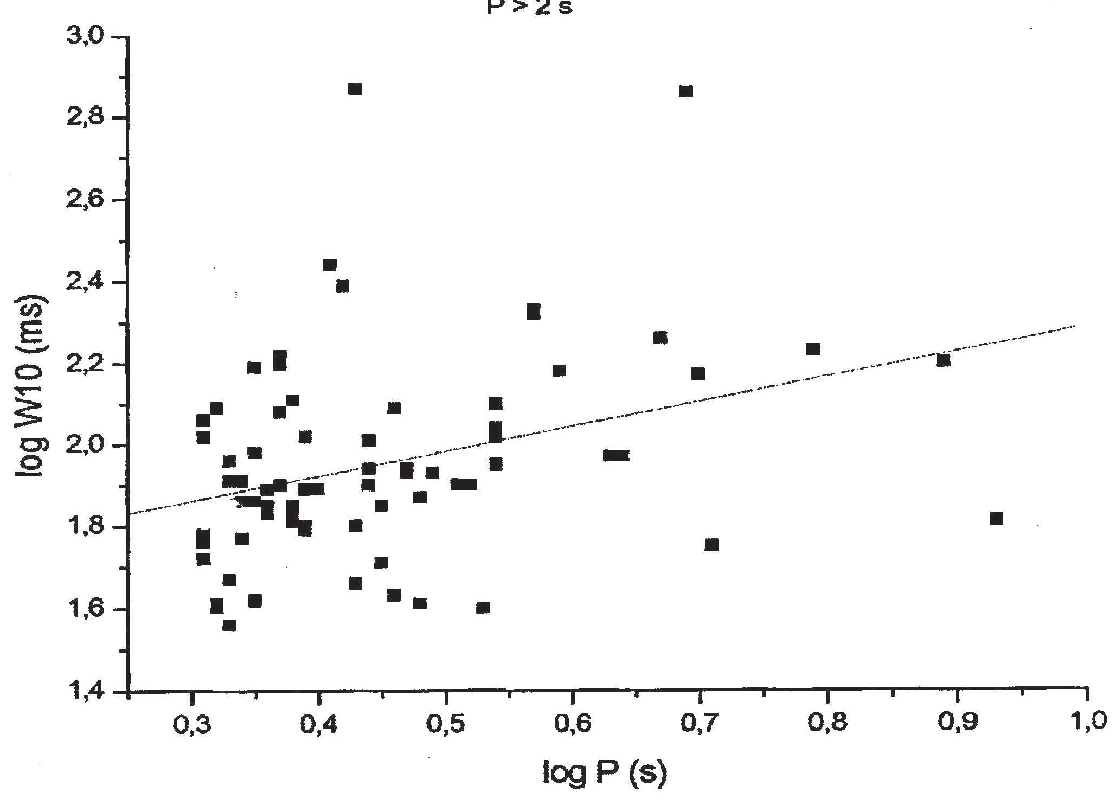}

Fig. 8 - The dependence of the pulse width on the period
\vspace{\baselineskip}

The quantiles of the distributions in Fig. 5-7 exceed 3, and therefore, they differ with a probability greater than 0.999.

 As for the pulse width for “normal” pulsars in which the pulse is formed at moderate distances significantly smaller than the radius of the light cylinder from the surface, it can be represented by the following expression:

\begin{equation}
W_{10} = P(r/r_{LC})^{1/2}=\left(\frac{2\pi r P}{c} \right) ^{1/2}
 \label{eq:ref}
\end{equation}

Thus, the pulse width (in the temporal measure) should increase with increasing period. The ATNF catalog data is indeed quite consistent with this dependence (Fig. 8):

\begin{equation}
log W_{10} = (0.61 \pm 0.21)log P + 1.68 \pm 0.10
 \label{eq:ref}
\end{equation}

with the correlation coefficient K = 0.33 and a random distribution probability p = 0.004.

It is more interesting to estimate  the coefficient of conversion of the rotational energy into the observed luminosity. For this purpose, $R_{lum}$ values were taken from the ATNF catalog, which were calculated by multiplying the flux density at the frequency of 400 MHz by the square of the distance and given in the catalog in units of $mJy \times kpc^2$. In ~\cite{Malov2006}, the equation was obtained for converting this “monochromatic” luminosity into the integrated radio luminosity in erg/s:

\begin{equation}
log L = (1.03\pm 0.03) log R_{lum} + (26.46 \pm 0.07)
 \label{eq:ref}
\end{equation}

The distributions of the luminosities calculated in this way are shown in Fig. 9. The average value <log L> = 28.42 for pulsars with 0.1 sec < P < 2 sec turns out to be several times higher than in pulsars with longer periods (<log L> = 27.66). The Gaussians inscribed in these distributions can be described by the following equations:

\begin{equation}
N = 128 exp \left[ - \left( \frac{log L - 28.55}{1.73} \right) ^2 \right] \qquad  (0.1 sec < P < 2 sec)
 \label{eq:ref}
\end{equation}
\begin{equation}
N = 10 exp \left[ - \left( \frac{log L - 28.16}{1.32} \right) ^2\right] \qquad \qquad \qquad   (P > 2 sec) 
 \label{eq:ref}
\end{equation}

However, according to the Kolmogorov-Smirnov test, these distributions are similar with a probability of more than 0.9.

Figure 10 shows the distribution of the transformation coefficient (efficiency).

\begin{equation}
\eta = \frac{L}{dE/dt}  
 \label{eq:ref}
\end{equation}

for two studied samples. In pulsars with longer periods, the transformation of the rotation energy into radio emission is more effective (the average values are log $\eta$ = -4.0 and -3.2, respectively).

Thus, the comparison of the parameters of the two populations under consideration leads to the conclusions that pulsars with P > 2 sec are decelerated more rapidly, the role of magnetic dipole braking is probably small in them, they have apparently lower radio luminosities and more efficient transfer of rotation energy to the observed radio emission.

\includegraphics[width=10cm, angle=0.8]{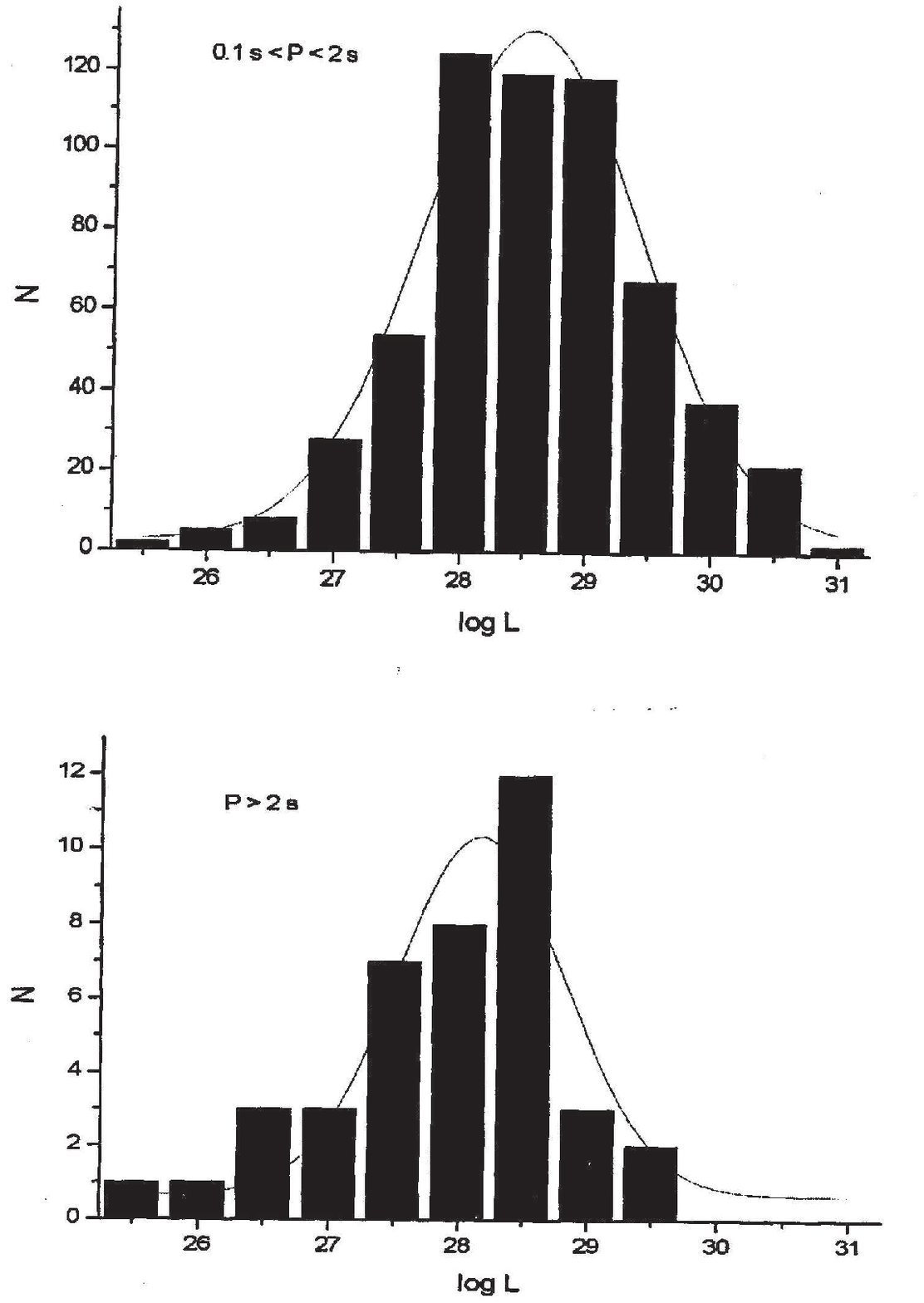}

Fig. 9 - Luminosity distributions for two groups of pulsars

\includegraphics[width=8cm]{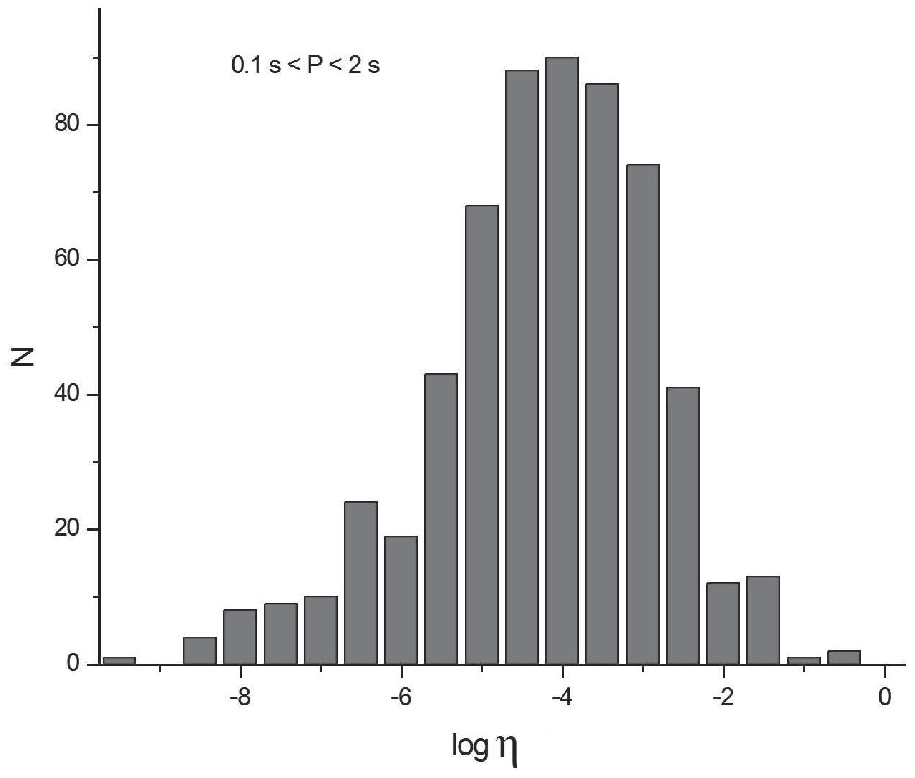}

\includegraphics[width=8cm]{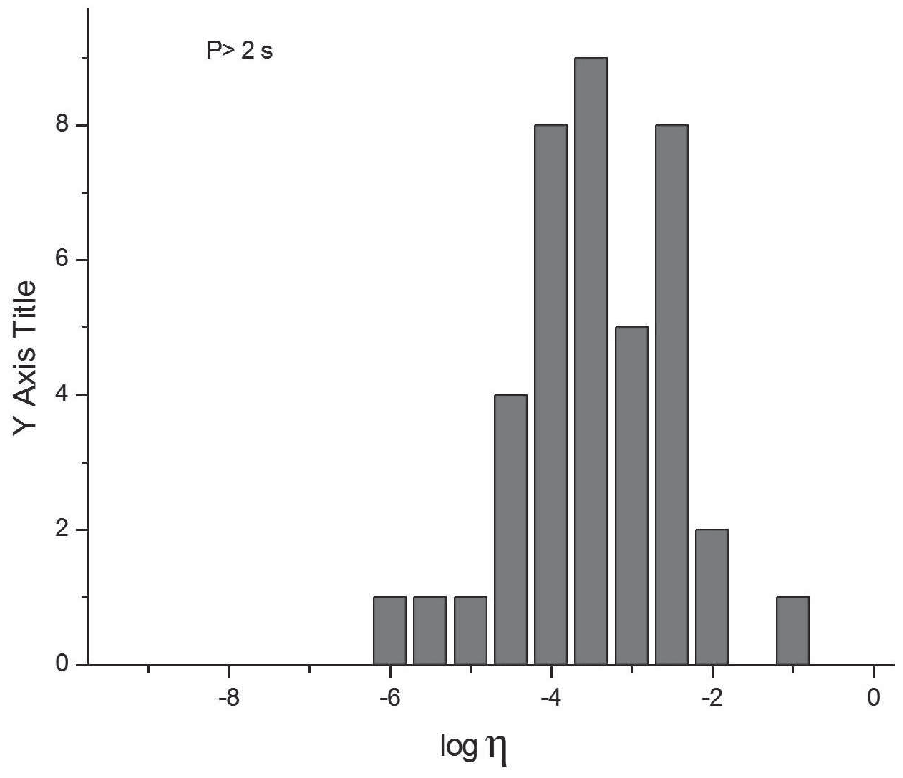}

Fig. 10 - Distributions of the transformation coefficient of the rotational energy into radio emission for two pulsar groups

\section{Comparison of pulsar parameters with magnetic fields more and less than 2.2 $\times$ 10$^{13}$ G}

Let’s now analyze the difference between pulsars with large values of dipole magnetic fields on the surface ($B_s > 2.2 \times 10^{13}$ G) from the rest of radio pulsars. We exclude from the analysis, as before, pulsars in globular clusters and in binary systems. The resulting sample is more than half (60\%) overlaps with the population considered in section 2 (with P > 2 sec). However, the parameters for this group are different from the parameters discussed in the previous section. Figure 11 shows the distribution of the periods of pulsars with magnetic fields above $2.2 \times 10^{13}$ G. The average period value for this group is <P> = 2.90 sec, which is several times higher than the value at the maximum of the period distribution  for the general population of normal pulsars ($\sim$ 0.6 sec) ~\cite{Loginov2014}.

\includegraphics[width=9.5cm]{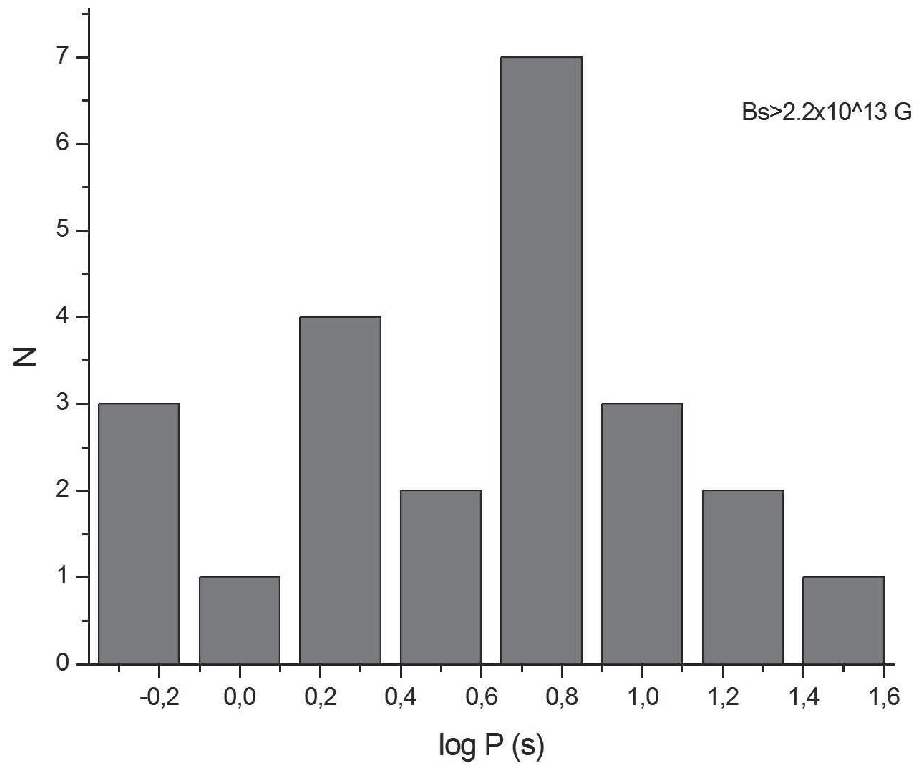}

Fig. 11 - Period distribution for 23 pulsars with large magnetic fields on the surface

\vspace{\baselineskip}

Figure 12 shows the distribution of period derivatives for the same sample.

The average value of <log dP/dt> = -12.29 is two to three orders of magnitude higher than that of the main bulk of pulsars, for which (excluding millisecond pulsars) the characteristic value is $\sim 10^{-15}$. But it is most interesting that the dependence dP/dt (P) (Fig. 13), in contrast to the corresponding dependence of the previous section (equation (5)), has the slope of the inscribed straight line in full agreement with the assumption of magnetic dipole braking ($dP/dt \propto P^{-1}$):

\begin{equation}
log (dP/dt) = (-1.13  \pm  0.14) logP - 11.82  \pm 0.09 ,
 \label{eq:ref}
\end{equation}

with the very high value of the correlation coefficient K = - 0.87 and the probability of a random distribution of less than $10^{-4}$. For these objects we can calculate magnetic induction using expression (1).

\includegraphics[width=10cm]{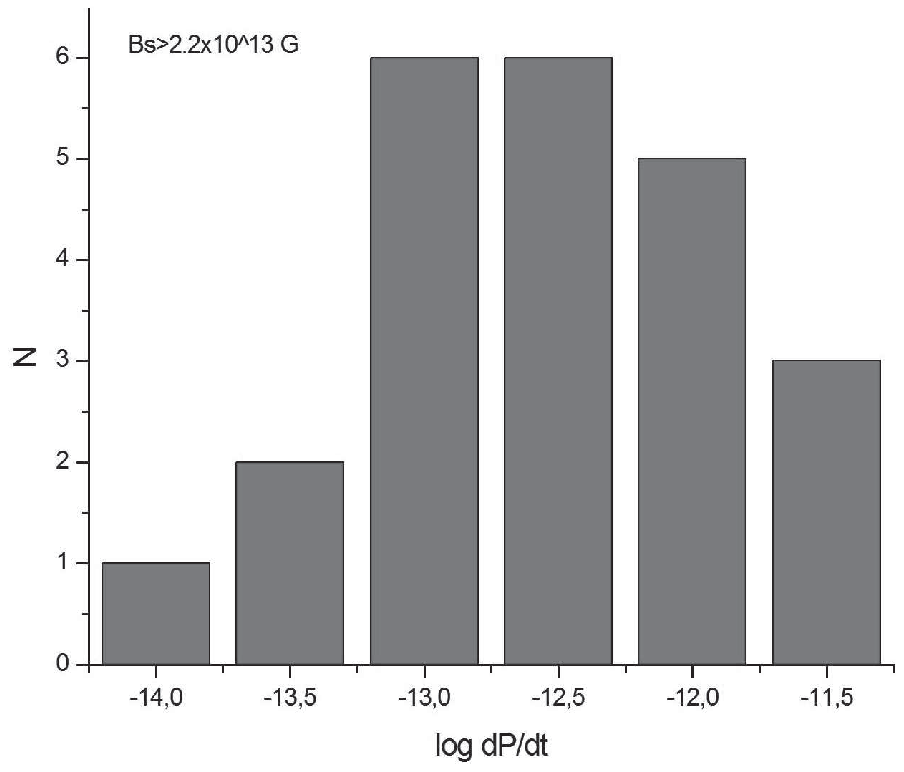}

Fig. 12 - Distribution of derivatives of the period for 23 pulsars with large magnetic fields
\vspace{\baselineskip}

\includegraphics[width=9.5cm]{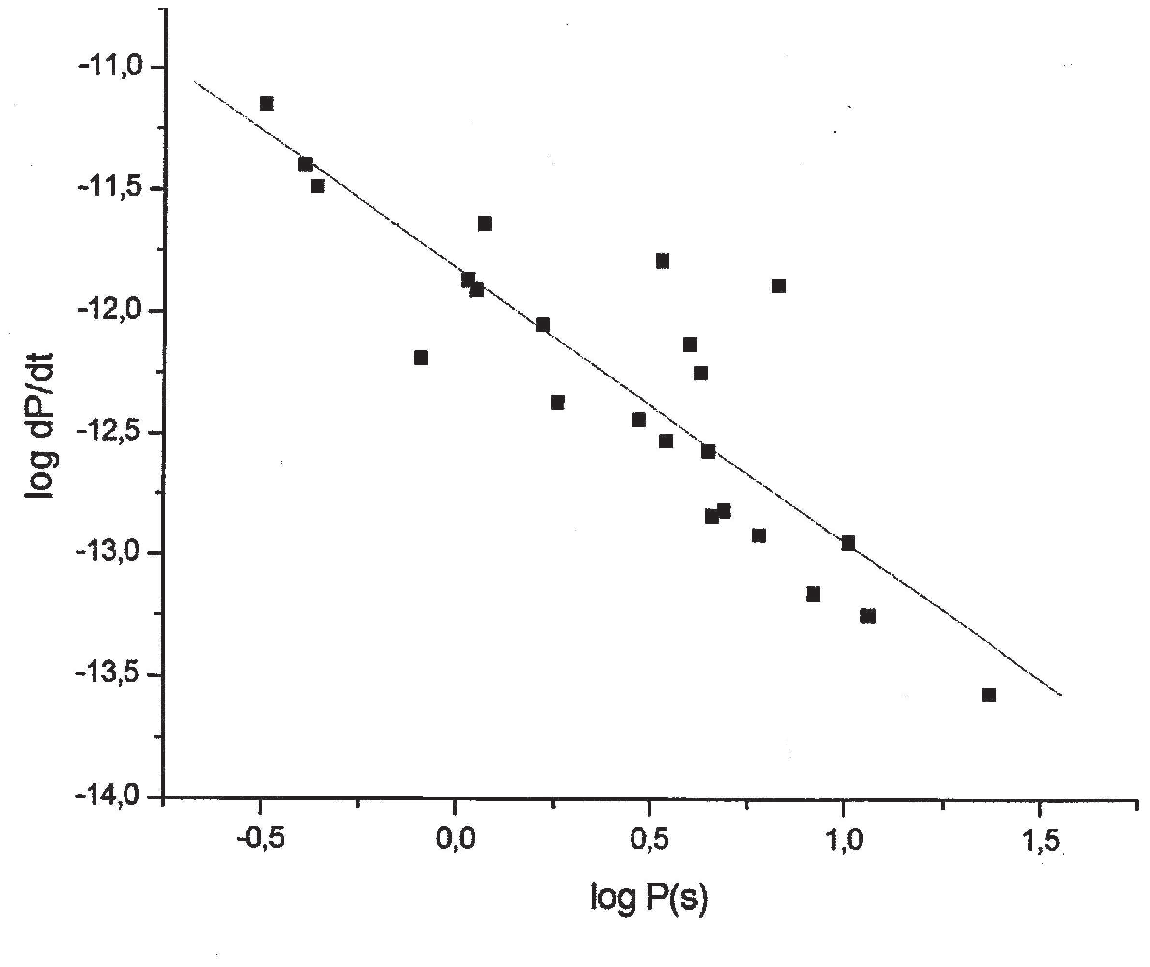}

Fig. 13 - The dependence of the derivative of the period on the period for pulsars with $B_s > 2.2 \times 10^{13}$ G
\vspace{\baselineskip}

The comparison of the magnetic fields on the light cylinder (Fig. 14) shows that, despite the large fields on the surface of the neutron star, the distribution of N(B$_{LC}$) for these objects coincides with the distribution for the main population with a probability of 80\%. The distributions of rates   of rotational energy losses  (Fig. 15)  have also  no noticeable difference.

Thus, pulsars with high magnetic fields on the surface are characterized by a more rapid braking of a neutron star, and this deceleration can be caused mainly by the loss of the angular momentum by magnetic dipole radiation. In addition, we note a lower value of the magnetic fields on the light cylinder, as in the sample of pulsars with P > 2 sec.

\includegraphics[width=9.5cm]{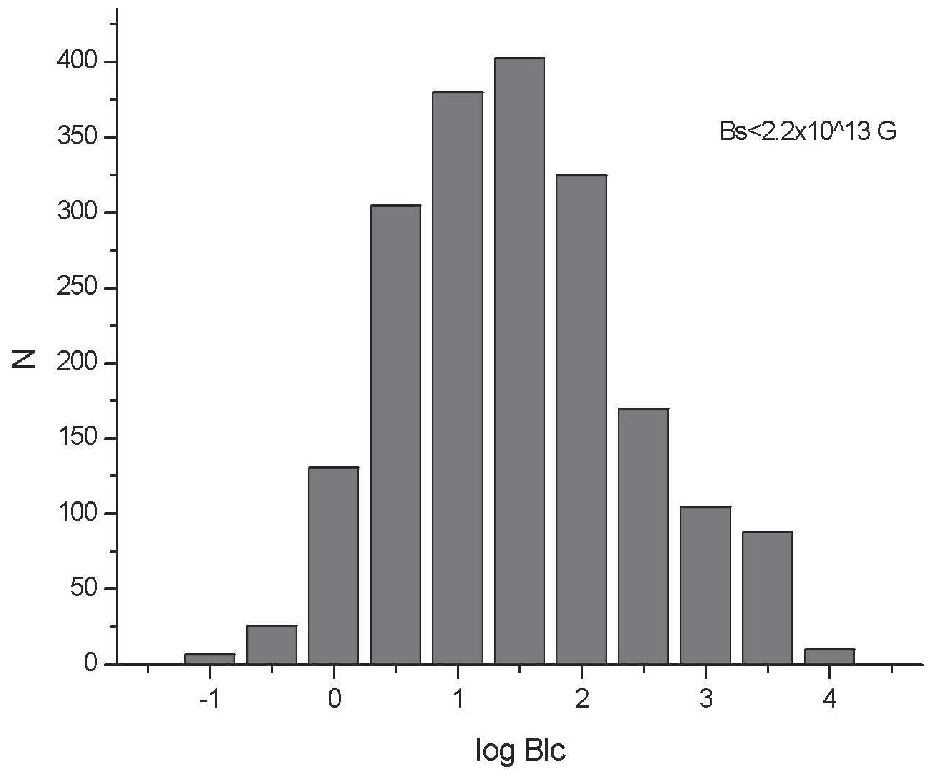}

\includegraphics[width=9.5cm]{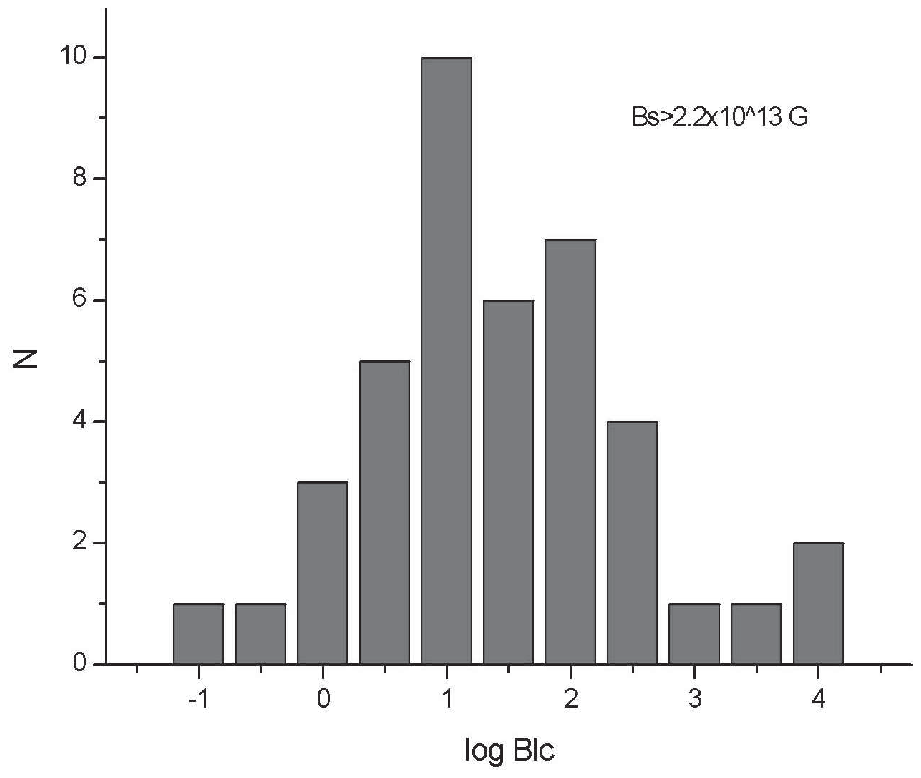}

Fig. 14 - Magnetic fields on the light cylinder for pulsars with large and normal magnetic fields

\includegraphics[width=10cm]{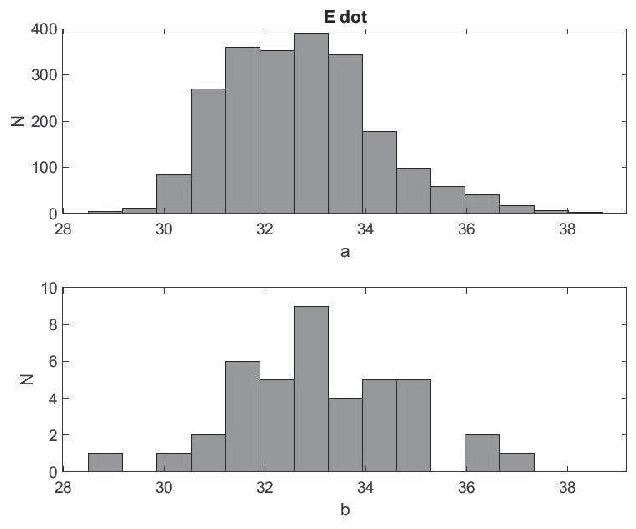}

Fig. 15 - Rotational energy loss rates for the main population (a) and pulsars with large magnetic fields on the surface (b)
\vspace{\baselineskip}

The additional and somewhat unexpected difference has been detected  between pulsars with large magnetic fields and pulsars with P > 2 sec  when comparing their ages (Fig. 16) and distributions along the Z coordinate (in height above the Galactic plane). The average age of pulsars with  periods longer than 2 seconds is almost two orders of magnitude higher  than the age of pulsars with large magnetic fields. For 155 pulsars with P > 2 sec, <log $\tau$> = 6.80, and for 23 pulsars with $B_s > 2.2 \times 10^{13}$ G <log $\tau$> = 4.97. The  equation of the Gaussian for pulsars with P > 2 s can be represented as:

\begin{equation}
N = 47 exp \left[ - \left( \frac{log \tau -7.10}{1.13} \right) ^2\right]
 \label{eq:ref}
\end{equation}

The location of the second population inside the Galactic disk is the additional evidence of their small ages. With the exception of one source with Z = 0.23 kpc, all the others are no higher than 100 pc (16 out of 23 pulsars have hights no more than 20 pc). Another picture is observed when analyzing long-period pulsars (Fig. 17). The average distance from the Galaxy plane is 440 pс for  pulsars with P > 2 sec. Moreover, when constructing Fig. 17 and calculating the mean value of Z, pulsars with very large Z were excluded: J0011 + 08 (Z = -4.34), J0928 + 06 (Z = 15.25), J1846-7403 (Z = -10.82), J2105 + 07 (Z = -10.59) and J2228-65 (Z= 17.92). Taking into account these objects gives <Z> = 0.74 kpc. The exponential curve  in Fig. 17 is described by the expression:

\begin{equation}
N = 252 exp\left[ -Z/0.27\right]
 \label{eq:ref}
\end{equation}

\includegraphics[width=10cm]{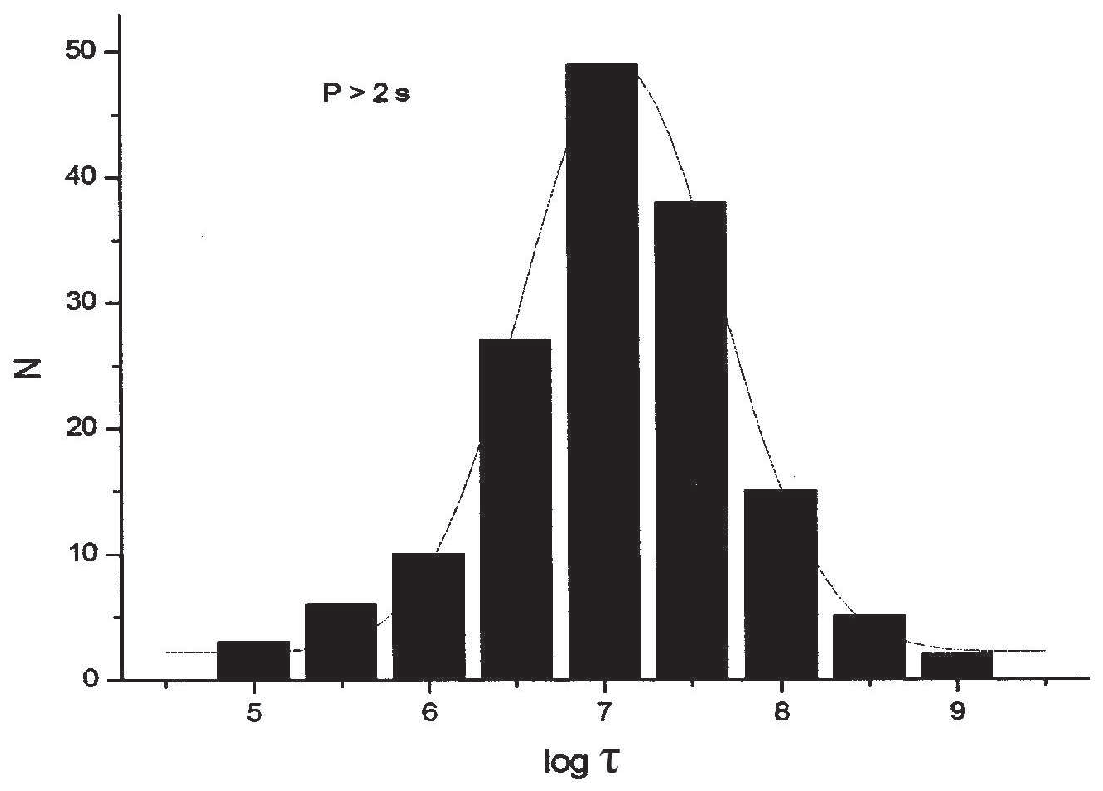}

\includegraphics[width=10cm]{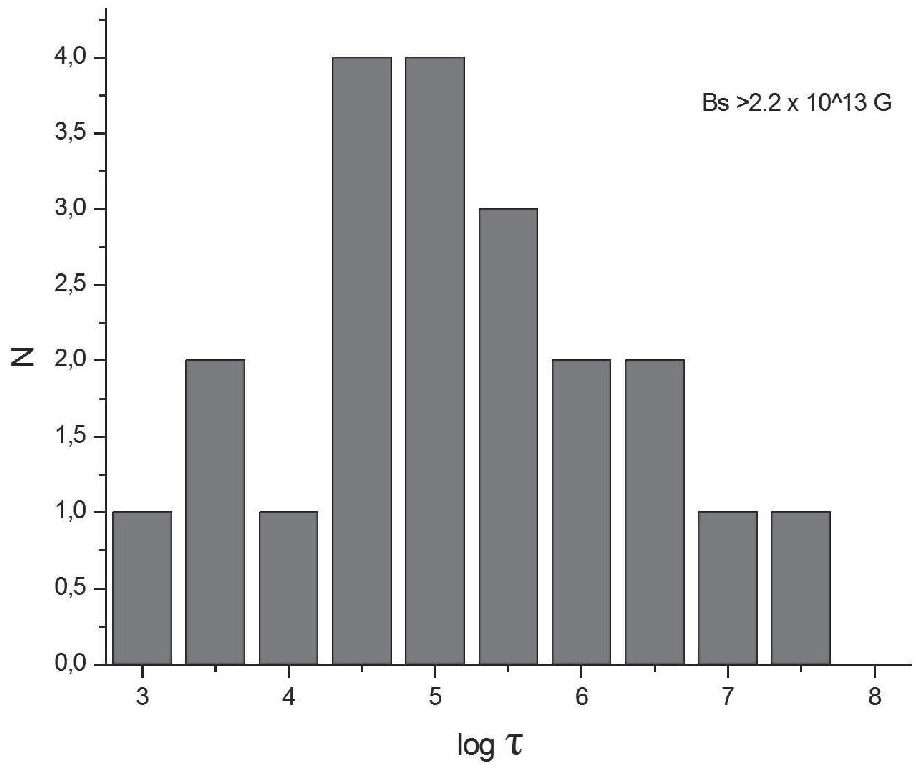}

Fig. 16 - Distributions of the characteristic age of pulsars with long periods (above) and large magnetic fields (below)

\includegraphics[width=10cm]{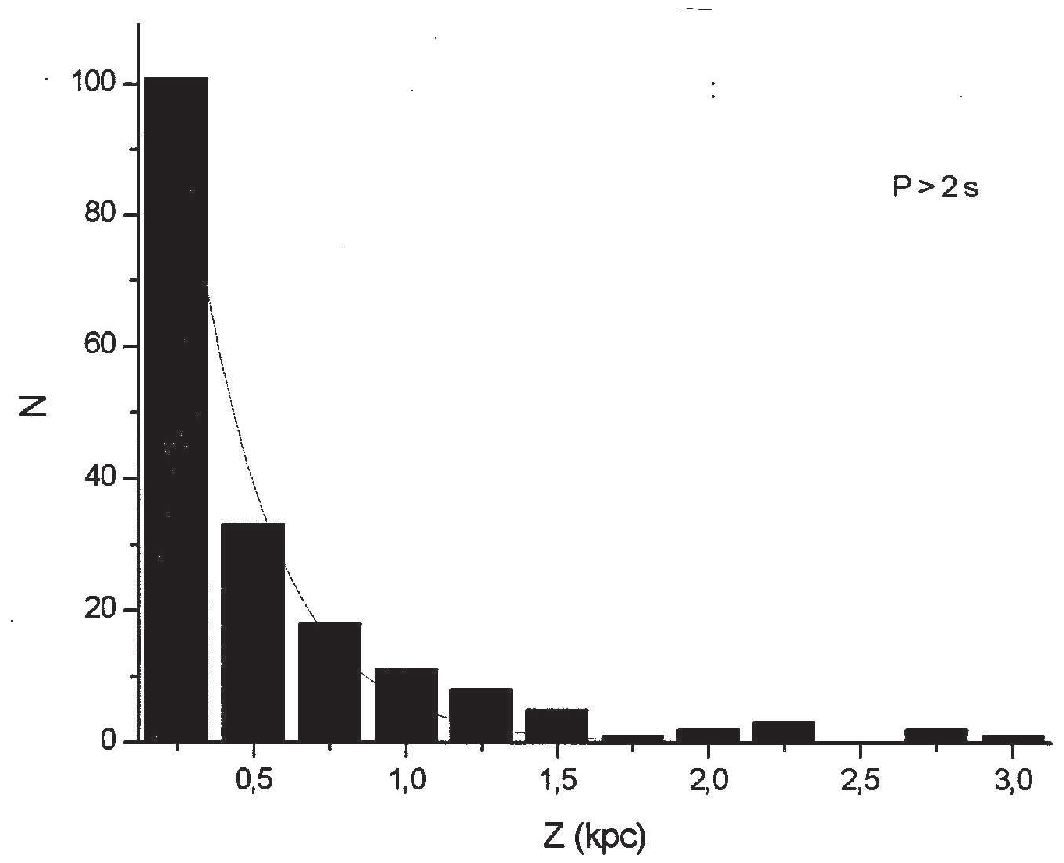}

Fig. 17 - The distribution of long-period pulsars in height above the Galactic plane

\section{Comparison of parameters of the considered pulsars and of AXP/SGR}

Let’s now compare the peculiarities of the pulsars discussed in the previous two sections with the characteristics of neutron stars exhibiting the flare activity in the X-ray and gamma ranges, anomalous X-ray pulsars (AXR) and soft gamma repeaters (SGR).

First of all, we note that these objects  are characterized by large periods (the average value according to the catalog ~\cite{mcgill} is 6.73 sec) (Fig. 18) and high values of the derivatives of the period (<log dP/dt> = -11.23) (Fig.19). However, the mean value of the rate of rotation energy losses <dЕ/dt> = 10$^{33}$ erg/s (Fig. 20) does not differ from the average values of dE/dt for the two previous samples. The magnetic fields on the light cylinder  turn out, on average, also  to be of the same order as for the bulk of the pulsars (Fig. 21).

\includegraphics[width=9.5cm]{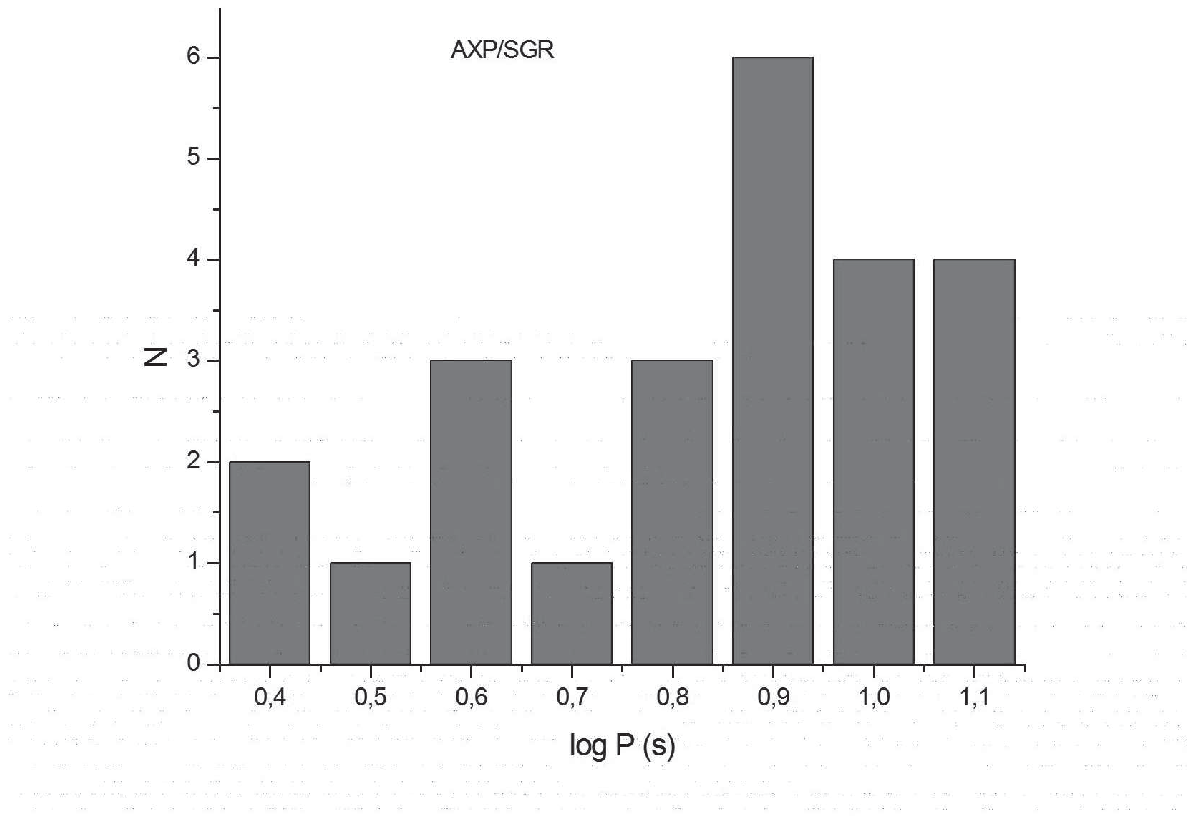}

Fig. 18 - Distribution of periods for AXP/SGR

\includegraphics[width=9.5cm]{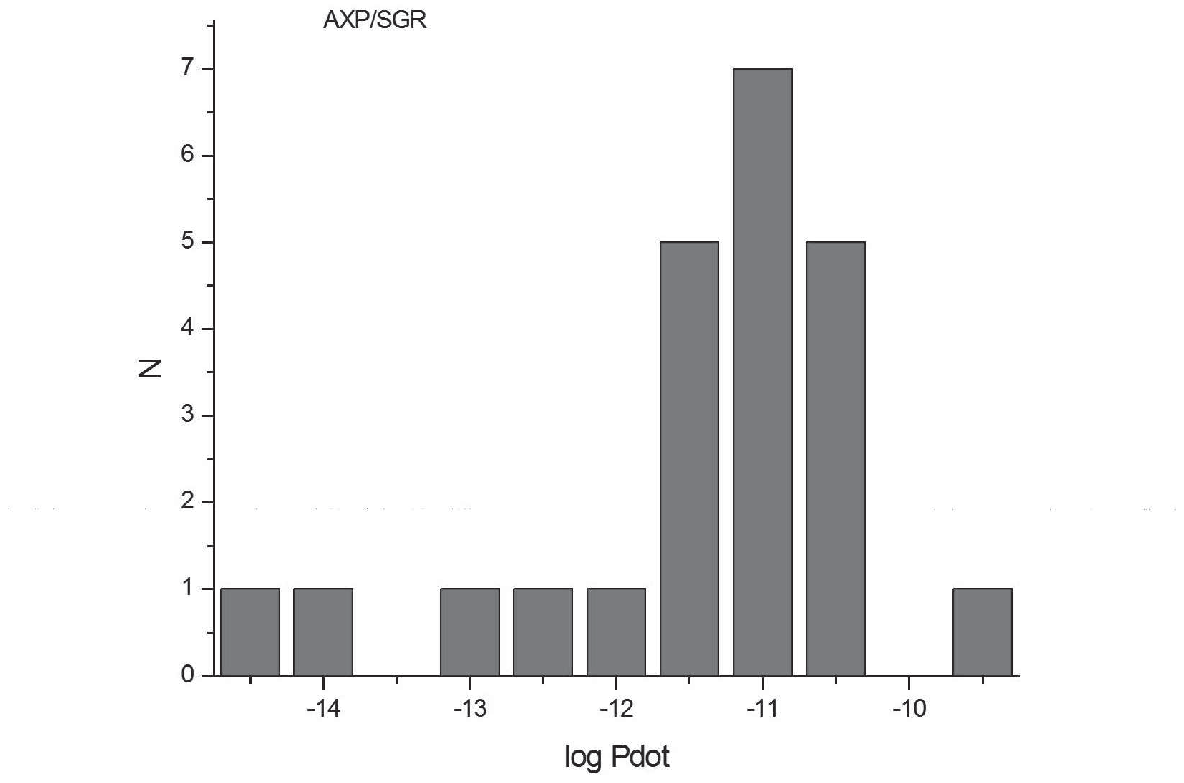}

Fig. 19 - Derivatives of periods for 23 AXP/SGR
\vspace{\baselineskip}

It is important to emphasize that the dependence dP/dt (P) (Fig. 22) is consistent with a similar dependence for pulsars with large magnetic fields and the opposite for pulsars with P > 2 sec:

\begin{equation}
log (dP/dt)  = (-1.77 \pm 1.10) log P - 9.85  \pm  0.89
 \label{eq:ref}
\end{equation}

Within the error limits, this equation is consistent with the predictions of the magnetic dipole model ($dP/dt \propto P^{-1}$), but less significant than for normal pulsars with $B_s > 2.2 \times 10^{13}$ G. The probability of random distribution is 12$\%$.

\includegraphics[width=9.5cm]{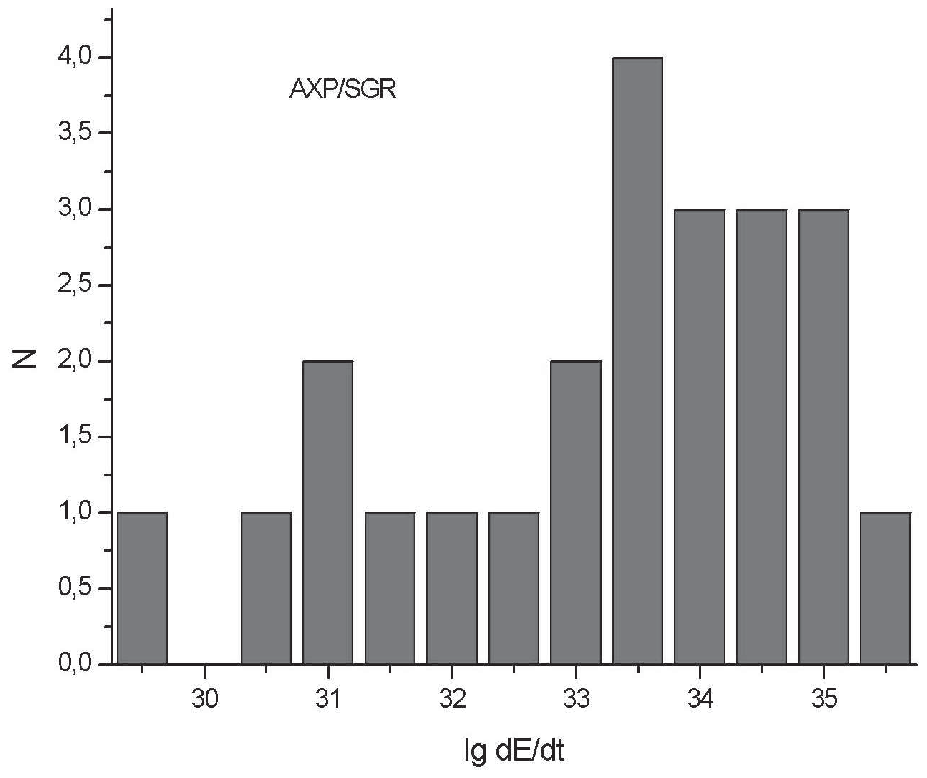}

Fig. 20 - Rates of rotational energy losses of AXP/SGR

\includegraphics[width=9.5cm]{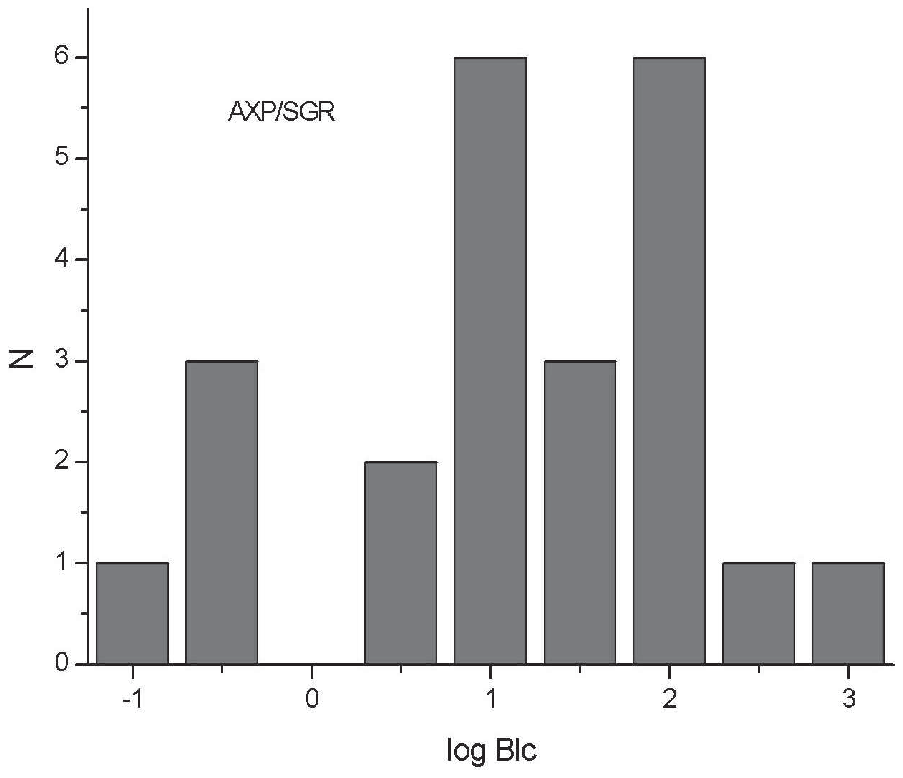}

Fig. 21 - Magnetic field distributions on the AXP/SGR light cylinders
\vspace{\baselineskip}

The proximity of AXP/SGR in properties to normal pulsars with large magnetic fields is also observed in their location in the Galactic disk. SGR 0418 + 5729 is the farthest from the plane (180 pc only) ~\cite{mcgill}. These populations are also close in age.

As evidenced by the presence of some AXP/SGR in supernova remnants for which estimates of their ages exist, these sources are not older than tens of thousands of years. The largest age $\sim$ 100 thousand years is indicated in the catalog [16] for the  Swift J1834.9-0846, which is possibly associated with SNR W41.

Thus, among normal pulsars, pulsars with $B_s > 2.2 \times 10^{13}$ G are similar to AXP/SGR in terms of their characteristics and the dependences between them.

\includegraphics[width=9.5cm]{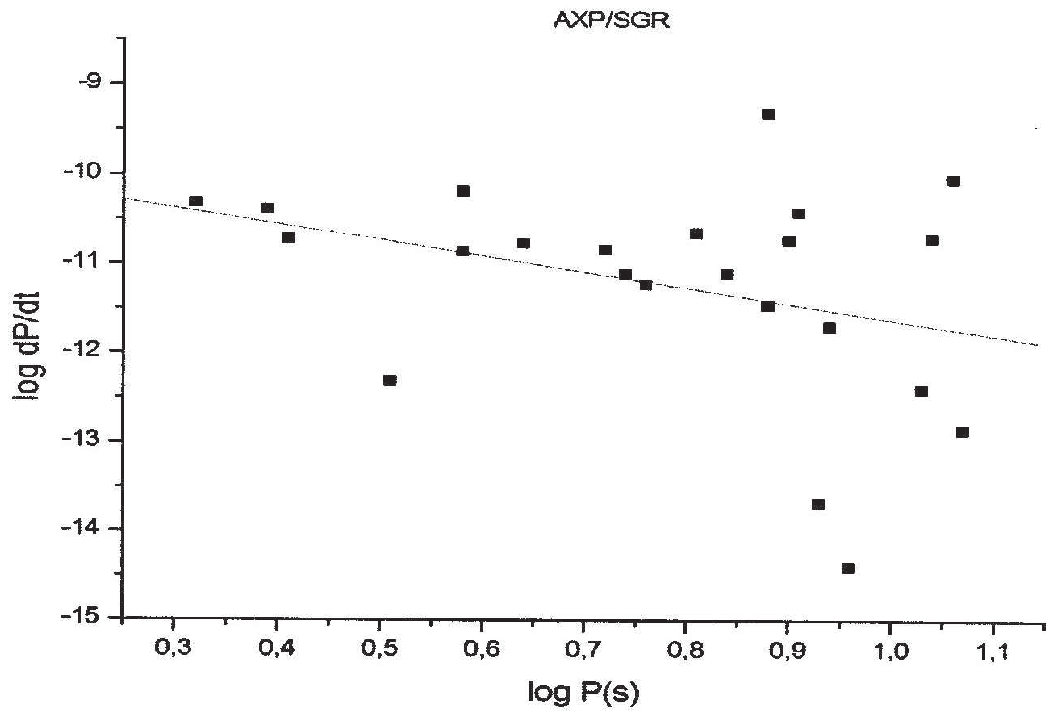}

Fig. 22 - Dependence of the period derivatives on the period for AXP/SGR

\section{Conclusions}

The comparative analysis of the parameters of radio pulsars with large magnetic fields on the surface of a neutron star ($B_s > 2.2 \times 10^{13}$ G) and pulsars with long periods (P > 2 sec) is performed. In these two populations, the fundamental difference is observed in the dependence of the derivative of the period on the period itself, which indicates that, if pulsars with large magnetic fields can slow down by the mechanism of magnetic dipole braking, then in long-period pulsars it is not applicable and other braking mechanisms must be used. Also, these two groups of radio pulsars differ in age and distribution relative to the plane of the Galaxy. Pulsars with P > 2 sec are, on average, approximately two orders of magnitude older and locate at larger heights above the plane. The characteristic parameters of AXP/SGR are close to the parameters of pulsars with $B_s > 2.2 \times 10^{13}$ G. The results obtained indicate the validity of the basic ideas about the evolution of radio pulsars. They are born mainly in the disk of the Galaxy, but, having high velocities, eventually leave it. Moreover, their periods increase due to losses of rotation energy and long-term objects must be accumulated at large distances from the disk. A detailed analysis of the possible causes of their differences in the observed manifestations of flare activity is planned for the next separate work.

\end{document}